\documentclass[aps,prl,preprint]{revtex4-1}
\usepackage[T1]{fontenc}
\usepackage[utf8]{inputenc} 
\usepackage[dvips]{graphics}
\usepackage{amssymb}
\usepackage{amsmath}
\usepackage{amsthm}
\usepackage{amsfonts}
\usepackage{subfigure}
\usepackage[]{color}
\usepackage{mathptmx}
\usepackage{epstopdf}

\newmuskip\pFqmuskip
\newcommand*\pFq[6][8]{%
	\begingroup 
	\pFqmuskip=#1mu\relax
	\mathcode`\,=\string"8000
	\begingroup\lccode`\~=`\,
	\lowercase{\endgroup\let~}\pFqcomma
	{}_{#2}F_{#3}{\left[\genfrac..{0pt}{}{#4}{#5};#6\right]}%
	\endgroup
}
\newcommand{\pFqcomma}{\mskip\pFqmuskip}

\newcommand{\bra}[1]{\ensuremath{\left\langle\, #1\,\right|}}
\newcommand{\ket}[1]{\ensuremath{\left|\,#1\,\right\rangle}}

\newcommand{\Tr}[1]{ \mbox{Tr}\left\{ #1 \right\} }

\begin{document}
	\title{A Complementary Resource Relation of Concurrence and Roughness for a two Qubits State}
	\author{Mauricio Reis} 
	\email{mreis@ufsj.edu.br}
	\address{Departamento de F\'isica e Matem\'atica, Universidade Federal de S\~ao
		Jo\~ao Del Rei, C.P. 131, Ouro Branco, MG, 36420 000, Brazil }
	\author{Adelcio C. Oliveira} 
	\email{adelcio@ufsj.edu.br}
	\address{Departamento de F\'isica e Matem\'atica, Universidade Federal de S\~ao
		Jo\~ao Del Rei, C.P. 131, Ouro Branco, MG, 36420 000, Brazil }
	\begin{abstract}
		Quantum resources lie at the core of quantum computation as they are responsible for the computational advantage in many tasks. The complementary relations are a pathway to understanding how the physical quantities are related. Here it was shown that there is a complementary relationship that evolves quantumness of a two qubits state and the degree of entanglement, respectively measured by Roughness and Concurrence. It was shown that there is a $\mathbb{R}^2$ surface that characterizes these quantities and can be understood as a projection of the Bloch Sphere. The distribution of states on the surface shows a concentration in the region of lower roughness and Concurrence, i.e., in the subset of states of greater entropy.  
	\end{abstract}
	
	\maketitle
	
	Since its advent as a physical theory, Quantum Mechanics brought
	several new insights into how nature works in the microscopic domain. In addition to the impact on understanding the physics of microscopic objects such as particles, and fields, Quantum Mechanics is responsible for the next leap in computing, quantum computers. They will not only be faster or with greater storage capacity but will allow us to address problems that are inaccessible by any other previously known computational method. Some of these exploits are already a reality, such as quantum cryptography \cite{Yin}, generation of random numbers \cite{Jacak}, and quantum teleportation \cite{Llewellyn}. 
	Recently, Google has claimed that they have obtained the quantum supremacy \cite{QSUP}, a group of researchers and Yin \textit{et al}
	carried out a cryptography experiment based on entanglement over a distance of 1,120 km \cite{Yin}. All of these advances in quantum technology have as a primary ingredient the use of quantum properties as a 
	resource \cite{Brandao} and are directly connected with the Quantum Contextuality \cite{AmaralPRL}. Although there is an established connection between quantum resources and non-locality, it is also possible to obtain them from a prepare-and-measure experiment approach, which explores the quantum nature of the state \cite{Amaral} as well practical application on  metrology \cite{Wenchao}. 
	The classicality depends on many factors, on the chosen observables, on how the measurement is made, and on the criterion adopted, see \cite{Oliveira2014} and references therein. To measure the degree of entanglement between two qubits, the Concurrence ($C$) is a good measure, it is zero when systems are not entangled and one for systems that are maximally entangled \cite{Wootters}.
	Along with Concurrence, other functions were proposed in order to measure and 
	identify if features in a given
	physical system are best described by Quantum Mechanics.
	Some of those {\em Quantumness} quantifiers make use of the statistical properties
	in the Wigner representation of a state to spot differences with the classical
	description {\em via} probability distribution. An example of a largely used function for this purpose is the Negativity function \cite{NEG}. Using a generalization of its ideas though, 
	a new quantity called {\em Roughness} were proposed \cite{Lemos2018}. 
	As one of desirable properties in a quantumness measure, 
	Roughness has its values limited, being zero for classical states
	and becoming one in the limit of extreme quantum states. 
	In a previous work \cite{Reis2018} prior to this one, a numerical investigation corroborated a mathematical relationship between Concurrence and Roughness. 
	Here it is shown that for a general two qubits states $\rho$ there is a complementary relation 
	
	\begin{equation}
	R_+^2(\rho) + \tilde{C}^2(\rho)=
	\tilde{N_e}(\rho) + \tilde{f_C}(\rho) - \kappa,
	\label{SURF}
	\end{equation}
	
	where $R_+$ is the sum of squared Roughness of each subsystem,
	$\tilde{C}$ is the Concurrence 
	re-scaled by a constant factor, $\tilde{N_e}$ is related to the total excitation number, $\tilde{f_C}$ is an auto-correlating function and $\kappa$ is a constant, more details in the following.  The right side of \ref{SURF} defines a $\mathbb{R}^2$ surface, that can be understood as a projection of the two qubits Bloch Sphere \cite{Remy} in the $(R_+,\tilde{C})$ set, $RC$. The main advantage is that the axis variables can be considered such as quantum resources. This is similar to the complementarity relation for a bipartite system exposed to an interferometer \cite{Jakob}, and its generalization for open systems \cite{Karen,JGJ}. Fan and collaborators \cite{Fan} have shown a complementary relation for quantum coherence and quantum correlations. The Roughness is sensible to quantum coherence and also to any other quantum aspect of the state.  Also, some proprieties of the $RC$ set were investigated thru a graphic approach.
	
	\section{Measuring the Quantumness of a Physical System}
	
	\textbf{Entanglement.} Quantum correlations between parts of a larger system 
	are often referred to as entanglement between its parts. 
	As a measure of entanglement of formation between parts of a 2-level quantum system represented by
	a density matrix $\rho$,
	Wooters and Hill \cite{Wootters} proposed the function known as Concurrence, which can be obtained as the non-negative eigenvalues of a ``flipped'' form of $\rho$ given by:
	$\tilde{\rho}=(\sigma_{y}\otimes\sigma_{y})\rho^*(\sigma_{y}\otimes\sigma_{y}),$
	where $\sigma_{y}$ is the matrix representation of Pauli spin operator and $\rho *$ is the complex conjugate of $\rho$.
	To obtain the Concurrence, one consider the operator 
	$\rho_F = \sqrt{\sqrt{\rho}\tilde{\rho}\sqrt{\rho}}$.
	Writting the eigenvalues of $\rho_F$ in decreasing order
	as$\{\lambda_{k}\}_{k=1}^4$, 
	the Concurrence of $\rho$ can then be formally written as:
	$C(\rho)=\mbox{max}(0,\lambda_{1}-\lambda_{2}-\lambda_{3}-\lambda_{4})$.
	Another useful function to be used for measuring entanglement
	of a bipartite state $\rho$ is the linear entropy, which can be obtained as:
	$\delta = 1-\Tr{\rho_1^2}\, ,$
	where $\rho_1$ stands for the local state obtained by tracing over the
	other part from the global state, $\rho_1 = \Tr{\rho}$. For pure global states, the two quantities are related:
	$C(\rho) = \sqrt{2\delta(\rho_1)} = \sqrt{2\delta(\rho_2)}$.
	
	The entanglement cost under operations preserving the positivity (PPT) of partial transpose can be measured by the negativity, although  PPT entanglement cost has some correlation with the entanglement of
	formation.  Miranowicz and Grudka have shown that concurrence and negativity can give different orderings of two-qubit states \cite{Miranowicz}. 
	
	\textbf{Roughness.} The Roughness of a quantum system were defined by Lemos {\em et al}
	to be the overall distance between the Wigner and Husimi representation
	of the system. For the $h_4$ algebra of bosonic systems, 
	it can be expressed by the formula:
	\begin{equation} 
	\label{eq:R}
	R(\Psi) = \sqrt{ 2\pi \int_{\Re^2} \!\!\!\! dq \, dp\, \big|
		W_{\Psi} (q,p) - Q_{\Psi} (q,p) \big| ^{2} } \, ,
	\end{equation}
	where $W_{\Psi} (q,p)$ and $Q_{\Psi} (q,p)$ are the Wigner
	and Husimi functions for the quantum state $\Psi$ according to the 
	position $q$ and momentum $p$ phase space quadratures.
	
	
	Following this definition, the Roughness of a density state
	is a real-valued function restricted to the $[0, 1]$ interval.
	Analytic expressions for the Roughness of traditionally physical
	systems with well known quantum features can be found in
	\cite{Lemos2018}.
	Departing from the Negativity's behaviour, Roughness is able to capture
	quantum features in states represented by for strictly positive
	Wigner functions. For example, physical systems represented
	by squeezed gaussian states, which present well known quantum effects,
	tend to have large $R$, since $|W-Q|$ increases with the squeezing.
	
	
	\section{A Complementary Relation between Roughness and Entanglement for a 2-qubit system}
	
	The proposal herein is to treat the
	quantum features of the general 2 qubit states represented by 
	a density operator $\rho$ in two  parts: one which
	could be measured performing local operations in each part and other  related
	to the quantum correlations between the parts measured by the 
	linear entropy $\delta(\rho)$. The first relations
	will be get in view of a {\em single} qubit state $\rho_1$,
	written in the $\{\ket{0}, \ket{1}\}$ basis as:
	\begin{equation}
	\rho_1 = \sum_{n,n^{\prime }\in\{0,1\}}A_{n,n^{\prime
	}}\left|\,n\,\right\rangle \left\langle\, n^{\prime }\,\right|
	\label{rho_1qb}
	\end{equation}
	with $\sum |A_{n,n^{\prime }}|^2 =1$ and $A_{n,n^{\prime }} = A_{n^{\prime
		},n}^*$. Using  this 2 conditions it is possible to represent this state
	operator  as a dimension 3 vector given by:
	\begin{equation}
	v =
	\begin{pmatrix}
	A_{00} \\
	\frac{1}{\sqrt{2}}A_{01} \\
	A_{11}%
	\end{pmatrix}%
	\, .
	\end{equation}
	Within this notation, the linear entropy of $\rho$ can be obtained by a
	product:
	\begin{equation}
	\delta(\rho) = 1 - \mbox{Tr}\left\{ \rho^2 \right\} = 1 - v^{\dagger}v\, ,
	\end{equation}
	where $v^{\dagger}$ the transposed and complex conjugated 
	version of $v$. Also, under this notation, the square of Roughness can be  written
	as a quadratic form:
	\begin{equation}
	R^2 = v^\dagger\Lambda v\, ,
	\end{equation}
	where $\Lambda$ can be obtained by computing the corresponding  entries from
	Appendix:
	\begin{equation}
	\Lambda = \frac{1}{108}%
	\begin{pmatrix}
	18 & 0 & -21 \\
	0 & 39 & 0 \\
	-21 & 0 & 55%
	\end{pmatrix}%
	\, .
	\label{Lambda}
	\end{equation}
	Putting $\Lambda$ in diagonal form and performing the corresponding  change
	of basis in $v$ one can recast the relations previously stated  to find how 
	$R^2$ of a single qubit is related to its linear entropy:
	\begin{equation}
	R^2(\rho) = \frac{55}{108} -\frac{37}{108} A_{00} (2-A_{00}) - \frac{39}{108}
	\delta(\rho)\,.  \label{rough_delta_1d}
	\end{equation}
	
	
	Now, a general 2 qubit state $\rho$ can be written as:
	\begin{equation}
	\rho = \sum_{mm^{\prime}kk^{\prime}\in\{0,1\}} 
	\Psi_{mm^{\prime}kk^{\prime}}\ket{mk}\bra{m^{\prime}k^{\prime}}\, ,  \label{2qb}
	\end{equation}
	where the each $\ket{mm^{\prime}}$ is a short form for $\ket{m}\otimes\ket{m^{\prime}}$, the product
	state of two independent Fock states for each subsystem, and $\bra{kk^{\prime}}$ the
	corresponding linear functional.
	Consider now the physical properties  which 
	would be detected \emph{locally}, i.e., performing local measurements on
	each subsystem alone. Such operations can  be
	described using partial trace operation acting over $\rho$:
	\begin{equation}
	\rho_1 = \sum_{mm^{\prime}k\in\{0,1\}} \Psi_{mm^{\prime}kk}\ket{mk}\bra{m^{\prime}k}\, .  \label{ptrace}
	\end{equation}
	Equation \eqref{ptrace} defines a state of the form \eqref{rho_1qb} where 
	$A_{00} = \Psi_{0000} + \Psi_{0011}$, 
	$A_{01} = \Psi_{0100} + \Psi_{0111}$, 
	$A_{10} = \Psi_{1000} + \Psi_{1011}$ and  
	$A_{11} = \Psi_{1100} + \Psi_{1111}$.
	
	Changing the system of interest and applying the corresponding operations, 
	analogous coefficients for $\rho_2$ are to be found.
	The states represented by $\rho_1$ and $\rho_2$ are two mixed states in the same form
	of the general 1-qubit states given  by \eqref{rho_1qb}. It is a matter of
	straightforward
	calculation  to obtain the Roughness of such states and express them in
	terms  of the linear entropy:
	\begin{equation}
	R^2(\rho_i) = \frac{1}{108}\left[37 \mu_i + 55 - 39\delta(\rho_i)\right]  \label{R2rhoA}\,, \end{equation}
	where $\mu_1= (\Psi_{0000} + \Psi_{0011})(\Psi_{0000} +
	\Psi_{0011}-2)$ and  $\mu_2= (\Psi_{0000} + \Psi_{1100})(\Psi_{0000} +
	\Psi_{1100}-2)$.
	
	In view of the arguments given
	by  \cite{Lemos2018}, expression \eqref{R2rhoA} and its  counterpart for 
	$\rho_2$ represents a scale of quantum features of a 2 qubit  state which
	could be measured and detected \emph{locally}.  
	Combining expressions given by \eqref{R2rhoA} and its corresponding
	expression for $\rho_{2}$, the following equation can be written:
	\begin{widetext}
		\begin{equation}
		\frac{R^2(\rho_1) + R^2(\rho_2)}{2}+
		\frac{39}{216}\left[\delta(\rho_1)+\delta(\rho_2)\right]= \frac{37}{216}\left\{z^2+w^2-2(z+w)\right\}+\frac{55}{108}
		\label{combsum}
		\end{equation}%
	\end{widetext}
	where $z = \Psi_{0000}+\Psi_{0011}$ and $w = \Psi_{0000}+\Psi_{1100}$. 
	
	The equation \eqref{combsum} expresses a non-negative quantity depending only on the
	diagonal coefficients in \eqref{2qb}, a general 2 qubit
	system. Since only normalized states are being considere, $0\leq z \leq 1$ and
	$0\leq w \leq 1$, so the quantity enclosed by braces in 
	equation %
	\eqref{combsum} is bound to the interval $[-2,0]$, with its maximum being
	achieved when $z=w=0$.
	According to this, the 
	combined sum from the left hand side of \eqref{combsum} is bound to values in the range: 
	$[1/6,55/108]$. 
	
	The quantities on the right hand side can be interpreted 
	as measurable properties of the physical system. On the other hand, it is possible to estimate the entanglement from randomized measurements \cite{Imai}, thus is possible to infer the mean Roughness using equation \ref{combsum}.
	Considering two separate measures of the mean photon number, 
	one should  find:
	\begin{equation}
	N_e = \Tr{(\hat{n}_1+\hat{n}_2)\rho} = 2-z-w.
	\end{equation}
	Now, $z$ and $w$ are the coefficients of $\ket{0}\bra{0}$
	in each reduced state operator $\rho_{1,2}$ obtained via partial trace
	operation. Due to this, is convenient to define the auto-correlation function
	of the localized fundamental states as:
	\begin{equation}
	f_C(z, w) = \frac{z^2+w^2}{2}\,.
	\label{fC_def}
	\end{equation}
	Using the above definitions, expression \eqref{combsum} can be
	rewritten as:
	\begin{equation}
	\begin{split}
	 \frac{R^2(\rho_1) + R^2(\rho_2)}{2}+
	\frac{39}{216}\left[\delta(\rho_1)+\delta(\rho_2)\right]=\\
	\frac{37}{108}(N_e + f_C) - \frac{19}{108}\,.
	\label{R2DNefC}
	\end{split}
	\end{equation}
	
	
	Both terms in the left hand side of 
	\eqref{R2DNefC} are affected by quantum features in $\rho$,
	and they are being related to the number of 
	localized excitation in $\rho$ and the 
	self correlation function $f_C$
	in a 
	complementary way: lowering $N_e$ by increasing
	$z$ and $w$ leaves a residual quantum effect.
	This can be interpreted as a collective effect in the ground state projector
	of the reduced density operators obtained {\em via} partial trace
	operation over $\rho$.
	
	By the uncertainty principle in quantum field theory, 
	the more the position of a particle 
	is definite,  the less the momentum is known. From 
	a relativistic point of view, when the momentum is large, particles can be 
	created making their numbers uncertain. These particles interact, so the more 
	you know about the location the less you know about the interactions. The 
	result presented here is similar, although the number of particles is constant, 
	the interaction between them can be determined by the state and measured by 
	concurrence. Those which have a well-defined location of the excitation are 
	states with no interaction, whereas in states of maximum entanglement, the 
	excitation is not local.
	
	\subsection{The Globally-pure states case}
	If the global state $\rho$ is pure, the relationship between
	Concurrence and the Linear Entropy of the reduced states
	$C(\rho) = \sqrt{2\delta(\rho_1)} = \sqrt{2\delta(\rho_2)}$,
	can be used.
	Also, just for aesthetic and simplicity reasons it is worthwhile to define:
	\begin{equation}
	R_+^2(\rho) = \frac{R^2(\rho_1) + R^2(\rho_2)}{2}\\
	\end{equation}
	and 
	\begin{equation}
	\tilde{C}(\rho)  = \frac{39}{216}C(\rho).
	\end{equation}	
	Then equation \eqref{R2DNefC} becomes:
	\begin{equation}
	R_+^2(\rho) + \tilde{C}^2(\rho)= \tilde{N_e}(\rho) + \tilde{f_C}(\rho) - \kappa,
	\label{R2CNefC}
	\end{equation}
	where $\tilde{N_e}(\rho)= \frac{37}{108}N_e $,  $\tilde{f_C}(\rho)=\frac{37}{108} f_C(\rho)$ and $\kappa=\frac{19}{108}$.
	
\begin{figure}
	\includegraphics{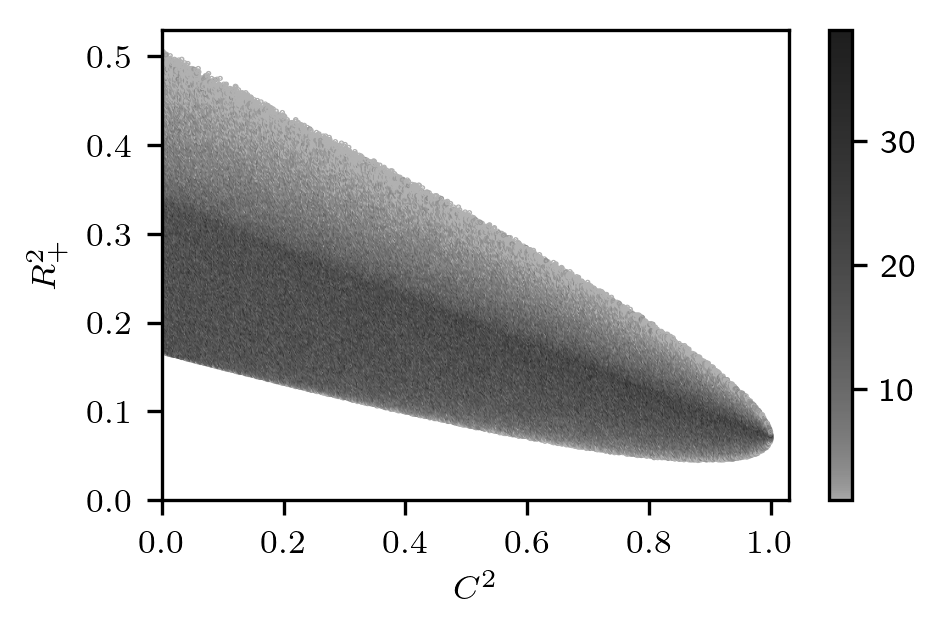}
	\caption{Numerical sampling of $1.57\times 10^{6}$ 2-qubit pure states. Gray (color online) scale indicates
	multiplicity of states in the vicinity of each point.}
	\label{pureFULL}
	\end{figure}

	Equation \eqref{R2CNefC} suggests the possibility of using 
	$R_+^2$ and $C^2$ to address complementary quantum features in the 
	2-qubit subspace. Figure \ref{pureFULL} shows a sampling of states considering $R^2$ and $C^2$ as parameters and illustrates the
	trade-off of local and global quantum features. The sampling were done using 
	the algorithm available in \cite{ginibre2009}, implemented by the Qutip Python package 
	\cite{Qutip1,Qutip2}.
	
	The map of possible pure 2-qubit states are restricted to a region in blade-like shape. At least 3 important points can be related to
	well known quantum states. At the tip of the blade, for $C^2=1$, lies the family of Bell states,
	$\ket{\Phi^{\pm}}=(\ket{00}\pm\ket{11})/\sqrt{2}$ and $\ket{\Psi^{\pm}}=(\ket{01}\pm\ket{10})/\sqrt{2}$.
	They all have the same $R_+^2$ value of $(31/432)\approx0.07176$.
	For states where $C^2=0$, the values of $R_+^2$ varies
	from a minimum of $1/6$ for $\rho=\ket{00}\bra{00}$ and the maximum
	of $55/108$ for $\rho=\ket{11}\bra{11}$. In this case, though 
	no entanglement between parts exists, it is still possible
	to have quantum features detected locally. For a better understanding
	of what those quantum features might mean, one could think about
	the nature of a pure vacuum state. Statistically, such a state
	is related to cooperative effects like Bose-Einstein condensates
	which exists at zero temperature. As for the product of
	two Fock states with a single excitation, the locally detectable 
	quantum feature can be related to having a pure state with a
	definite number of excitation. This can be compared with the sampling of
	non-pure states which follows next.
	
	\subsection{The Globally non-pure states case}
	
	\begin{figure}
		\includegraphics{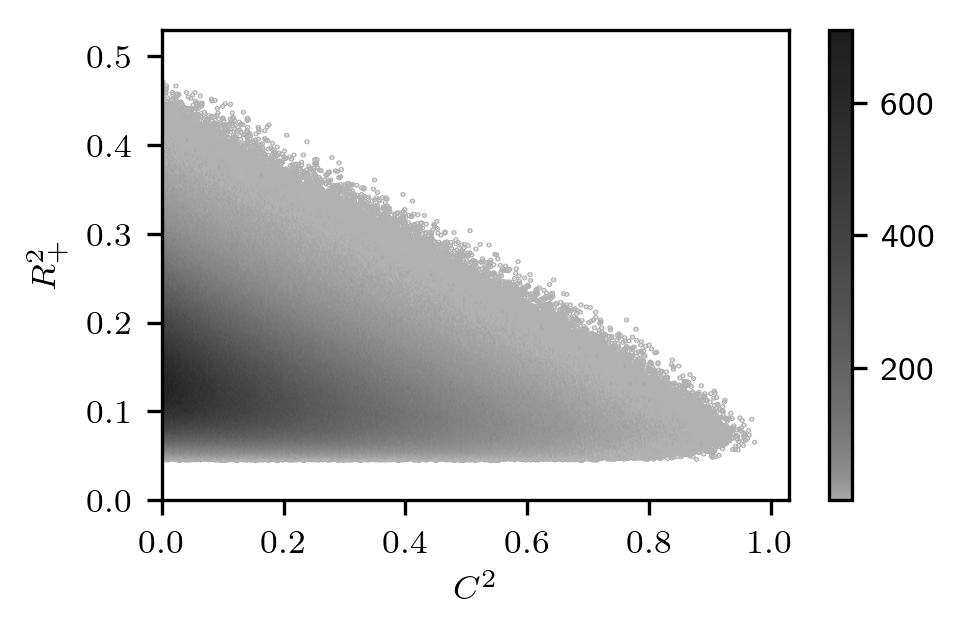}
		\caption{Numerical sampling of $1.26\times 10^{7}$ 2-qubit rank-2 states. Gray (color online) scale indicates multiplicity of states in the vicinity of each point.}
		\label{nonpureR2C2}
	\end{figure}
	
	Figure \ref{nonpureR2C2} shows the sampling in the 2-qubits subspace without
	requiring the global states to be pure. It was also done with
	the algorithm given in \cite{ginibre2009}, through the implementation of
	\cite{Qutip1}. The density matrix representing the global quantum state
	were expected to have rank 2. It means the states could have the structures like
	$\rho = p\ket{\psi_0}\bra{\psi_0}+(1-p)\ket{\psi_1}\bra{\psi_1}$ i.e., they could be
	written as a sum of up to 2 projectors. 
	Though pure states were still possible to be obtained, 
	they would have a negligible proportion in this subspace. Still, sampling in this
	way show how the loss of purity affects the quantum features mapped in 
	the $(C^2, R_+^2)$ parameter space. As can be seen in the figure, there is
	a concentration of states in the lower left corner, indicating the great
	majority of the states in this sampling have much lower quantum features,
	be it in the form of entanglement or local quantum features. But it 
	also indicates the existance of a large region where global states with similar quantum
	features analogous of those in the globally pure state cases. 
	Further numerical investigation indicates the states in the intercepting region
	of the two cases have higher purity.
	
	\textbf{Closing Marks.} The relationship of Roughness and linear entropy of the parts, in a 2-qubit system, were cast in a way all quantum correlation terms of the system were bound to be expressed through two other quantities with physical mean-ing which could be measured locally in each of the parts. It allows to take
	the two quantities as a complementary pair for describing quantum
	features of a physical system. Most of the world observed macroscopically appears
	to be absent of quantum features. 
	Yet, even when considering a single part system, quantum
	effects can still be a remarkable feature. And it is 
	a well known fact that the interaction of a physical system
	with a thermal bath leads to a decoherence process which makes 
	quantum  information to flow from system to environment 
	in an irreversible way. Quantum mechanics of open systems describe 
	such a process by the loss of state's purity and this is highlighted 
	by the linear entropy of the state.
	In a sense, physical systems represented by pure states
	are expected to have more quantum features than the non-pure
	states. Complementary to this, Roughness measure quantum features
	through the differences between Wigner and Husimi representations
	of the states. For globally pure states, those relationships leads
	to a bounded region in the $(C^2, R_+^2)$ parameter space. For the 
	non pure states of rank-2 density matrices the otherwise bound region
	were broken, with the great majority of states being found to have fewer
	quantum features. This is in line with general knowledge of 
	what is observed for macroscopic physical systems.
Low	Roughness and low Concurrence states are the ones obtained most frequently, while the region of high values of $R_+$ and $C$ are inaccessible for states described by \eqref{ptrace}. This corroborates the hypothesis that some states are more robust and form the pointer states, used in the construction of the Quantum Darwinism hypothesis \cite{Zurek}. Thus, Quantum Darwinism is not only the result of natural selection caused by decoherence, but it is also a consequence of the high density of states with low values of $R_+$ and $C$. This set of states is also the one with the highest entropy, since for non-pure states there is a strong correlation between Roughness and Shanon Entropy \cite{Lemos2018}.
	
	\textit{Acknowledgements:} The authors acknowledge the support of Brazilian agency Fundação de Amparo a Pesquisa do Estado de Minas Gerais through grant No. APQ-01366-16. 
	
	\subsection{Appendix}
	In order to calculate the coefficients of \eqref{Lambda}, it has been used the Appendix C of \cite{Lemos2018}.The terms of $R^2$ for a general state \ref{rho_1qb} the Roughness can be written as:
	\begin{widetext}
	\begin{equation}
	R^2(\rho) = \sum_{nm,n'm'}A_{nm}^*A_{n'm'}\left(
	R^2_{\Pi_{mn}\Pi_{m'n}}+R^2_{\Psi_{mn}\Psi_{m'n}}
	-R^2_{\Pi_{mn}\Psi_{m'n}}-R^2_{\Psi_{mn}\Pi_{m'n}}\right)
	\end{equation}
	\end{widetext}
	
	where the various terms within the brackets are results
	of the evaluation of integrals. According to the
	referenced article,it is:
	
	\begin{align}
	R^2_{\Pi_{mn}\Pi_{m'n'}}=&\delta_{nn'}\delta_{mm'}; \\
	R^2_{\Psi_{mn}\Psi_{m'n}}=&\frac{\delta_{n-m,n'-m'}}{\sqrt{n!m!n'!m'!}}
	\left(\frac{1}{2}\right)^{1+v}v!;
	\end{align}
	\begin{widetext}
		\begin{eqnarray}
		R^2_{\Pi_{mn}\Psi_{m'n'}}=\frac{2\delta_{n-m,n'-m'}}{3}(-1)^{Y}
		\sqrt{\frac{Y!}{X!Y!Y'!}} 2^{X-Y}\left(\frac{1}{3}\right)^u
		\pFq{2}{1}{-Y, u + 1}{X - Y + 1}{4/3};\\
		R^2_{\Pi_{mn}\Psi_{m'n'}}=\frac{2\delta_{n-m,n'-m'}}{3}(-1)^{Y'}
		\sqrt{\frac{Y'!}{X!Y!X'!}} 2^{X'-Y'}\left(\frac{1}{3}\right)^{u'}
		\pFq{2}{1}{-Y', u' + 1}{X' - Y' + 1}{4/3},
		\end{eqnarray}
	\end{widetext}
	where $X$ and $Y$ stands for max$(n,m)$ and min$(n,m)$, with the corresponding
	relations for the primed indexes, $u=(X - Y + X' + Y') / 2 $, $v=(n+m+n'+m')/2$ and 
	$_2F_1$ is the Generalized Hypergeometric Function using the 
	bracketed arguments. Evaluating the expressions above for
	the case of a density state $\rho$ as in \eqref{rho_1qb} leads to the following expression for the squared Roughness:
	
	\begin{equation}
	R^2(\rho) = \frac{55}{108} -\frac{37}{108} A_{00} (2-A_{00}) - \frac{39}{108}
	\delta(\rho)\,.  \label{rough_ap}
	\end{equation}

\end{document}